# Angular diameter estimation of interferometric calibrators
## Example of $\lambda$ Gruis, calibrator for VLTI-AMBER

P. Cruzalèbes[1], A. Jorissen[2], S. Sacuto[3] and D. Bonneau[1]

[1] UMR CNRS 6525 H. Fizeau, Univ. de Nice-Sophia Antipolis, Observatoire de la Côte d'Azur, Av. Copernic, F-06130 Grasse
[2] Institut d'Astronomie et d'Astrophysique, Univ. Libre de Bruxelles, Campus Plaine C.P. 226, Bd du Triomphe, B-1050 Bruxelles
[3] Institute of Astronomy, University of Vienna, Türkenschanzstrasse 17, A-1180 Wien

**ABSTRACT**

*Context.* Accurate long-baseline interferometric measurements require careful calibration with reference stars. Small calibrators with high angular diameter accuracy ensure the true visibility uncertainty to be dominated by the measurement errors.
*Aims.* We review some indirect methods for estimating angular diameter, using various types of input data. Each diameter estimate, obtained for the test-case calibrator star $\lambda$ Gru, is compared with the value 2.71 mas found in the Bordé calibrator catalogue published in 2002.
*Methods.* Angular size estimations from spectral type, spectral index, in-band magnitude, broadband photometry, and spectrophotometry give close estimates of the angular diameter, with slightly variable uncertainties. Fits on photometry and spectrophotometry need physical atmosphere models with "plausible" stellar parameters. Angular diameter uncertainties were estimated by means of residual bootstrapping confidence intervals. All numerical results and graphical outputs presented in this paper were obtained using the routines developed under PV-WAVE®, which compose the modular software suite SPIDAST, created to calibrate and interpret spectroscopic and interferometric measurements, particularly those obtained with VLTI-AMBER.
*Results.* The final angular diameter estimate 2.70 mas of $\lambda$ Gru, with 68% confidence interval 2.65-2.81 mas, is obtained by fit of the MARCS model on the ISO-SWS 2.38-27.5 $\mu$m spectrum, with the stellar parameters $T_e$ = 4 250 K, $\log g$ = 2.0, $z$ = 0.0 dex, $\mathcal{M}$ = 1.0 $\mathcal{M}_\odot$, and $\xi_t$ = 2.0 km s$^{-1}$.

## 1. Introduction

Recent improvements in the optical long-baseline interferometers need good knowledge of the calibrator fundamental parameters and of their brightness distribution. In our paper, we review different methods of angular diameter estimation for a test-case calibrator star and compare the results obtained with the corresponding value found in the calibrator catalogue usually considered as reference for optical interferometry.

In Sect. 2, we recall some basics of interferometric calibration and study in Sect. 3 the influence of the angular diameter uncertainty on the visibility, applied to the uniform-disk model case. In Sect. 4, we review the criteria to be fulfiled by a potential calibrator and introduce the calibrator star $\lambda$ Gru. In Sect. 5, we recall the distinction between the direct and the indirect approaches of angular diameter estimation and present various calibrator catalogues presently available for optical interferometry. In Sect. 6, we give the main characteristics of the most used stellar atmosphere models used for our study, particularly those of MARCS. In Sect. 7, we applied some methods of angular diameter estimation to the case of $\lambda$ Gru, based on : the Morgan-Keenan-Kellman spectral type (Sect. 7.1), the colour index (Sect. 7.2), the in-band magnitude (Sect. 7.3), the broadband photometry (Sect. 7.4), and the spectrophotometry (Sect. 7.5). In Sect. 8, we discuss the results in terms of diameter uncertainty (Sect. 8.1), of fundamental stellar parameters (Sect. 8.2), and of atmosphere model parameters (Sect. 8.3). We conclude in Sect. 9, and present the main functionalities of the software tool,

*Send offprint requests to*: pierre.cruzalebes@oca.eu

which we have developed in order to process, calibrate, and interpret the VLTI-AMBER measurements. The method used to compute the uncertainties is described in Appendix A, the dereddening process in Appendix B, and the residual bootstrap method in Appendix C.

## 2. Interferometric calibration

Absolute calibration of long-baseline spectro-interferometric observations of scientific targets, such as fluxes, visibilities, differential, and closure phases, needs simultaneous measurements with calibrator targets, allowing determination of the instrumental response during the observing run (Mozurkewich et al., 1991; Boden, 2003; van Belle & van Belle, 2005). The true (i.e. calibrated) visibility function is $V_{\rm true} = \mu_{\rm sci}/R_{\rm V}$, where $\mu_{\rm sci}$ is the measured visibility of the scientific target, and $R_{\rm V}$ the instrumental response (in visibility).

In principle, when we consider the instrument as a linear optical system, observation of a point-like calibrator gives the system response. Thus, the visibility response $R_{\rm V}$ is simply equal to the measured visibility of the calibrator $\mu_{\rm cal}$. Unfortunately, instrumental and atmospheric limitations make the instrument unstable and contribute to destroying this linearity. To get a reliable estimate of the instrumental response during the observing run, scientific and calibrator targets must be observed under similar conditions. With the VLTI-AMBER instrument described by Petrov et al. (2007), it has been showed that the estimator used to measure the fringe visibility also depends on the signal-to-noise ratio (Tatulli et al., 2007; Millour et al., 2008), so that cal-ibrators as bright as their corresponding scientific targets must be found. Most of the time, it is difficult to find unresolved and bright enough calibrators in directions close to a given bright scientific target. To determine the system response, it is preferable to use bright, but non-point-like, calibrators, observed under instrumental conditions similar to those of the scientific targets, rather than dimmer point-like sources observed under different conditions. The price to pay for this choice is the need for an independent estimation of the calibrator brightness distribution (Boden, 2007).

If the system response in visibility is given by $R_V = \mu_{cal}/V_{mod}$, where $V_{mod}$ is the calibrator model visibility, then the true visibility becomes $V_{true} = V_{mod}\, \mu_{sci}/\mu_{cal}$. Considering a calibrator with a circularly-symmetric brightness distribution, with angular diameter $\phi$, the model visibility function at the wavelength $\lambda$, for the sky-projected baselength $B$, is given by the normalized Hankel transform (of order 0) of the radial brightness distribution, according to the Van Cittert-Zernike theorem (Goodman, 1985)

$$V_{mod} = \frac{2\pi \left| \int_0^1 L_\lambda(r)\, J_0\left(\pi r \phi \frac{B}{\lambda}\right) r\, dr \right|}{M_\lambda}, \quad (1)$$

where $r$ is the distance from the star centre expressed in radius units ($r = 0$ in direction to the disk centre, $r = 1$ towards the limb), $J_0$ the zeroth-order Bessel function of the first kind, $L_\lambda$ the monochromatic brightness distribution, herafter called spectral radiance, i.e. the monochromatic emitted luminous intensity (in W m$^{-2}$ $\mu$m$^{-1}$ sr$^{-1}$), and $M_\lambda$ is the spectral radiant exitance, i.e. the monochromatic emitted luminous flux (in W m$^{-2}$ $\mu$m$^{-1}$), integration of the spectral radiance into the full solid angle of an hemisphere around the emitting area (Malacara & Thompson, 2001).

## 3. Effect of the diameter uncertainty

A bad knowledge of the calibrator angular diameter can skew the true visibility estimate. For a small angular-diameter uncertainty $\sigma_\phi$, the model-visibility absolute uncertainty $\sigma_{V_{mod}}$ is usually computed applying the approximation of the first-order Taylor series expansion of the visibility function, increasingly inaccurate for non-linear equations,

$$\sigma_{V_{mod}} \approx \left| \frac{\partial V_{mod}}{\partial \phi} \right| \sigma_\phi. \quad (2)$$

In the case of the uniform-disk (ud) model, the monochromatic visibility function is $V_{ud} = 2|J_1(x)|/x$, where $J_1$ is the first-order Bessel function of the first kind, and $x = \pi \phi B/\lambda$ is a dimensionless argument, which can be also expressed as

$$x = \frac{\pi^2}{648} \frac{B(m)}{\lambda(\mu m)} \phi(mas). \quad (3)$$

The first partial derivative of the visibility with respect to the angular diameter transforms Eq. (2) into

$$\sigma_{V_{ud}} \approx 2 |J_2(x)| \frac{\sigma_\phi}{\phi}, \quad (4)$$

where $J_2$ is the second-order Bessel function of the first kind.

The left-hand panel of Fig. 1 shows the variation in $V_{ud}$ against $x$, while the right-hand panel shows the variation in the ratio $\sigma_{V_{ud}}/(\sigma_\phi/\phi)$, deduced from the first-order expansion of the visibility. Evidence that the first-order approximation of the standard deviation is inaccurate can be found particularly at the inflexion points of the visibility function ($x \approx 5.136, 8.417, 11.620...$), where this ratio should not drop to zero. One can notice that the second-order Taylor expansion deduced from Eq. (A.2),

$$\sigma_{V_{ud}}^2 \approx 4 [J_2(x)]^2 \left(\frac{\sigma_\phi}{\phi}\right)^2 - [J_2(x) - x J_3(x)]^2 \left(\frac{\sigma_\phi}{\phi}\right)^4, \quad (5)$$

gives negative values of the variance at the same points, which is a clear indication that higher order Taylor expansions would be needed.

Knowing that the amplitude of the first maximum of $J_2(x)$ reaches 0.4865 for $x \approx 3.0542$ (Andrews, 1981), we can infer that the visibility uncertainty of the ud-model due to the calibrator diameter uncertainty never exceeds

$$\max(\sigma_{V_{ud}}) \approx 0.973 \frac{\sigma_\phi}{\phi} \approx \frac{\sigma_\phi}{\phi}, \quad (6)$$

a maximum value that only depends on the relative uncertainty of the angular diameter. It results that, if one wants to get an absolute uncertainty of the science true visibility $\sigma_{V_{true}} = \sqrt{\sigma_{V_{mod}}^2 + \sigma_\mu^2}$ dominated by a given measured visibility error $\sigma_\mu = \sqrt{\sigma_{\mu_{cal}}^2 + \sigma_{\mu_{sci}}^2}$ for any calibrator angular diameter, the relative precision of the estimation of this diameter $\sigma_\phi/\phi$ must be lower than $\sigma_\mu$. For example, calibrator angular diameters estimated with relative uncertainties lower than 1% ensure the science true visibilities to be dominated by experimental visibility errors greater than 0.01.

If the relative uncertainty of the model diameter is higher than the experimental visibility absolute error ($\sigma_\phi/\phi > \sigma_\mu$), one can still find values of the calibrator diameter for which the absolute uncertainty of the science true visibility is dominated by the measurement errors. This can be achieved by numerical inversion of Eq. (4), in finding the values of the argument $x$ corresponding to model visibility absolute uncertainties lower than a given value of the measurement error $\sigma_\mu$, for a given diameter relative uncertainty $\sigma_\phi/\phi$. Because of the quasi-periodic behaviour of $\sigma_{V_{ud}}$ against $x$, as shown in the right panel of Fig. 1, many sets of diameter values, enclosing each zero of the $\sigma_{V_{ud}}$ function, can fulfil the condition $\sigma_{V_{ud}} \leq \sigma_\mu$. Then, we can define the value $x_0$, below which any ud-calibrator diameter would contribute to the global visibility uncertainty less than the experimental errors, thanks to the inversion of the relation $|J_2(x_0)| = \eta/2$, where $\eta = \sigma_\mu/(\sigma_\phi/\phi)$. To obtain the corresponding value of the angular diameter $\phi_0$, we can use Eq. (3)

$$\phi_0(mas) \approx 65.6 \frac{\lambda(\mu m)}{B(m)} x_0. \quad (7)$$

For example, if the calibrator angular diameters are estimated with 10% relative uncertainties, while the experimental visibility errors are 0.01 (i.e. $\eta = 10$), a model visibility error lower than 0.01 can be obtained for $x \leq 0.6435$. Using Eq. (7), we find that this is achieved for any ud-calibrator smaller than 0.93 mas, with a 100-m baselength interferometer operating at 2.2 $\mu$m. Table 1 gives some typical values of $\phi_0$ under which $\sigma_{V_{ud}} < 0.01$, for various diameter relative uncertainties, with various baselengths, also at 2.2 $\mu$m.

With $B = 330$ m, $\lambda = 2.15$ $\mu$m, and $\sigma_\phi/\phi = 5\%$, we find that any calibrator smaller than 0.42 mas gives $\sigma_{V_{ud}} < 0.011$, i.e. $\sigma_{V_{ud}^2} = 2\, V_{ud}\, \sigma_{V_{ud}} < 0.02$, a result very close to that of van Belle & van Belle (2005), who give 0.45 mas for the same value of $\eta$. Figure 2 shows the baselength dependency of $\phi_0$ at wavelengths ofs 10 (upper line), 2.2, 1.25, and 0.7 $\mu$m (lower line) on a log-log scale, with $\sigma_\mu = 0.01$ and $\sigma_\phi/\phi = 10\%$.

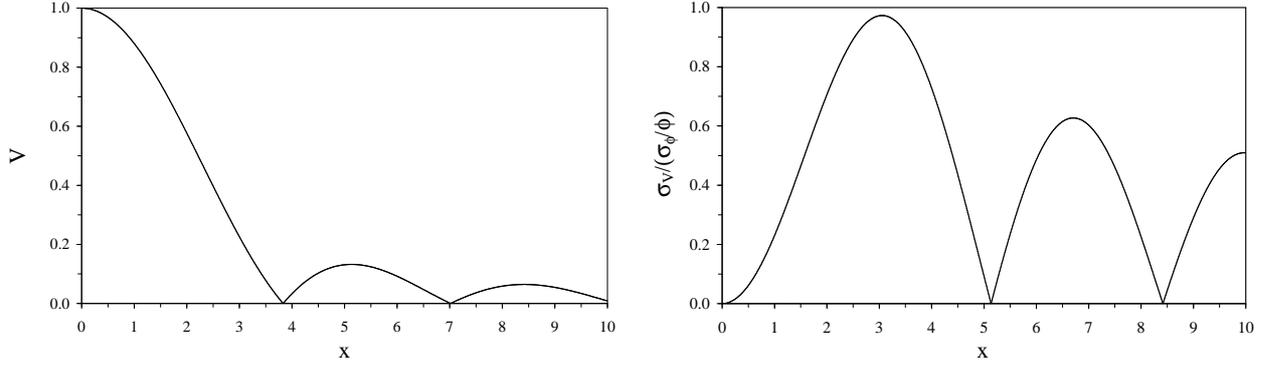

**Fig. 1.** Left panel: plot of the uniform-disk visibility function against the argument $x = \pi\phi B/\lambda$. Right panel: plot of the ratio of the model visibility uncertainty to the angular diameter relative uncertainty against $x$, given by the first-order Taylor series expansion of the visibility.

**Table 1.** Values of the angular diameter $\phi_0$ (in mas), under which $\sigma_{V_{ud}} < 0.01$, at $\lambda = 2.2\,\mu m$.

| $\sigma_\phi/\phi$ | B=20 m | B=50 m | B=100 m |
|---|---|---|---|
| 5% | 6.70 | 2.68 | 1.34 |
| 10% | 4.65 | 1.86 | 0.93 |
| 20% | 3.26 | 1.30 | 0.65 |

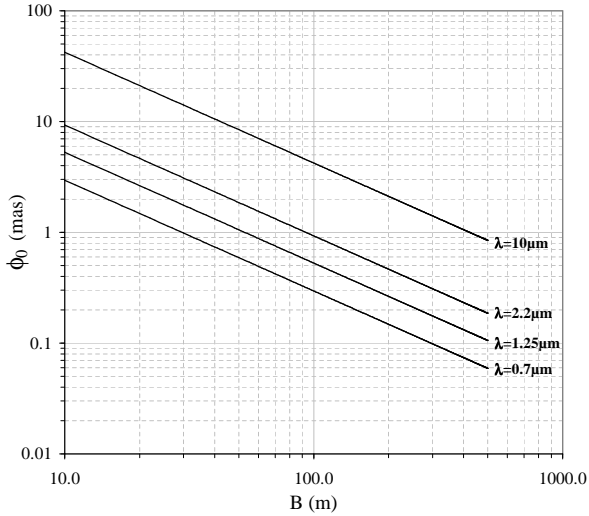

**Fig. 2.** Log-log plot of the angular diameter $\phi_0$ against the sky-projected baselength, for $\lambda = 10$ (upper line), 2.2, 1.25, and 0.7 $\mu$m (lower line), with $\sigma_\mu = 0.01$ and $\sigma_\phi/\phi = 10\%$. $\phi_0$ is such that the lack of precision in the size of any uniform-disk calibrator smaller than $\phi_0$ has no significant impact upon the final errors.

We can conclude from this short study that the choice of suitable ud-calibrators for long-baseline optical interferometry depends on the ratio $\eta$ of the absolute measurement error of the visibility $\sigma_\mu$ to the calibrator angular size prediction error $\sigma_\phi/\phi$. If $\eta \leq 1$, any ud-calibrator is suitable, i.e. contributes less than the measurement error to the global budget error in visibility, because of its angular diameter uncertainty. If $\eta > 1$, we find that any ud-calibrator with an angular diameter less than a value $\phi_0$, such as $|J_2(\pi\phi_0 B/\lambda)| = \eta/2$, is suitable.

## 4. Choosing the calibrators

Choosing calibration targets for a specific scientific programme in a given instrumental configuration is a critical point of the absolute calibration of interferometric measurements. If one wants to determine the visibility of scientific targets with a high degree of accuracy, not only the angular diameters of the calibrators need to be carefully estimated, but also their brightness distributions, which are known to deviate slightly from simple ud-profiles. This implies that suitable calibrators belong to well-known, intensively-studied, and easily-modelled object classes.

As said in Sect. 2, point-like calibrators as bright as their associated scientific targets are ideal interferometric calibrators, which are unfortunately rarely available. Partially resolved sources may also be considered as suitable calibrators, provided that their brightness distribution can be accurately modelled. This excludes irregular and rapid variables, evolved stars, or stellar objects embedded in a complex and varying circumstellar environment involving disks, shells, etc., which are potentially revealed by an infrared excess in the spectral energy distribution (SED).

Since we are concerned with studies of the circumstellar environment and of brightness asymmetries on the surface of evolved giants and supergiants, observed at high angular resolution with VLTI-AMBER in the near infrared (NIR), we consider as "good" calibrators the celestial targets fulfilling the following criteria :

1. small angular distance (< 10°) to the scientific targets,
2. spectral type not later than K, with luminosity class III at the most (no supergiant nor intrinsically bright giant),
3. NIR apparent magnitudes as close as possible to the scientific target ($\Delta m < 3$, i.e. flux ratio below 16),
4. angular diameter as small as possible but at least smaller than the scientific target,
5. no near-infrared excess observed in the spectral energy distribution (SED discrepancy with a blackbody radiator within ±1% in the NIR domain),
6. no evidence for variability identified in the CDS-SIMBAD database[1],
7. preferably source unicity, possibly multiplicity with far (> 2″) and/or faint ($\Delta m > 5$) companion(s) not seen in the observation field of the instrument, thus not affecting interferometric measurements,
8. no evidence for non centro-symmetric geometry.

---
[1] simbad.u-strasbg.fr/simbad/

To choose calibrators during the observation preparation phase, we usually cross-compare the output lists given by many calibrator selector tools: JMMC-SearchCal[2] (Bonneau et al., 2006), MSC-getCal[3] (NASA Exoplanet Science Institute, 2008), and ESO-CalVin[4]. The 2MASS catalogue[5] (Skrutskie et al., 2006) gives the NIR magnitudes. The infrared SEDs are extracted from the ISO-SWS[6] database of spectra (Leech et al., 2003), or the IRAS-LRS[7] database (Volk & Cohen, 1989), if no ISO-SWS spectrum is available. If multiplicity is suspected, the associated parameters can be found in the CCDM Catalogue (Dommanget & Nys, 2002).

The present paper uses the reference giant star $\lambda$ Gru (HD 209688) as test case, selected to calibrate the interferometric measurements of the scientific target $\pi_1$ Gru (HD 212087), that we observed in Oct. 2007 with the VLTI-AMBER instrument. The target $\lambda$ Gru has been used several times as calibrator for interferometry (Di Folco et al., 2004; Kervella et al., 2004; Wittkowski et al., 2006). The following set of basic information can be found for this star:

- equatorial coordinates (J2000): $\alpha$ = 22 h 06 m 06.885 s, and $\delta$ = $-39\degree\,32'\,36.07''$
- galactic coordinates (J2000): $l$ = 2.2153°, and $b$ = $-53.6743\degree$
- parallax: 13.20(78) mas (Perryman et al., 1997)
- spectral type: initially classified as M3III by Buscombe (1962), then as K3III since Houk (1978)
- apparent broadband magnitudes gathered in Table 2
- infrared fluxes: $f_{12\mu m}$ = 11.71(59), $f_{25\mu m}$ = 2.95(18), and $f_{60\mu m}$ = 0.41(6) from IRAS (in Jy) (Beichman et al., 1988)
- infrared spectrophotometry: from 2.38 to 45.21 $\mu$m with ISO-SWS01 (Sloan et al., 2003)
- angular diameter: 2.71(3) mas (limb-darkened) in the catalogue of calibrator stars for LBSI[8] (Bordé et al., 2002), revised to 2.75(3) mas by Di Folco et al. (2004) from observations with VLTI/VINCI.

Note the use of a concise notation for the uncertainties, e.g. 2.71(3) for the angular diameter, instead of the standard notation, e.g. 2.71±0.03. It must be understood that the number in parentheses in the concise notation is the numerical value of the standard uncertainty referred to the corresponding last digits of the quoted result. By extension, a value like $2.71^{+0.04}_{-0.02}$ in the standard notation becomes $2.71\binom{4}{2}$ in the concise notation. Unless otherwise stated, we use the concise notation to report the uncertainties throughout the present paper. Let us also add that symmetric and nonsymmetric uncertainties are computed from confidence intervals with 68% confidence level, corresponding to 1$\sigma$ errors for the normal distribution, as stated in Appendix A. Figures 3 and 4 respectively show the broadband absolute photometry and the ISO-SWS spectrophotometry of $\lambda$ Gru deduced from the magnitude and flux measurements, compared with a 4 250-K blackbody radiator.

---

[2] www.jmmc.fr/searchcal_page.htm
[3] nexsciweb.ipac.caltech.edu/gcWeb/gcWeb.jsp
[4] www.eso.org/observing/etc/
[5] www.ipac.caltech.edu/2mass/
[6] irsa.ipac.caltech.edu/data/SWS/
[7] www.iras.ucalgary.ca/satellites/Iras/getlrs.html
[8] cdsarc.u-strasbg.fr/viz-bin/Cat?J/A+A/393/183

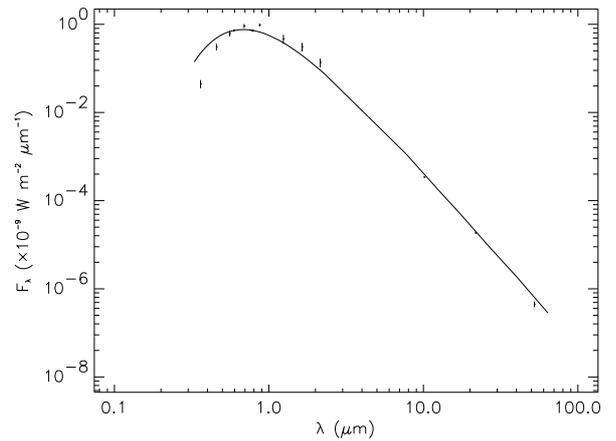

**Fig. 3.** Log-log plot of the broadband absolute photometry (in $10^{-9}$ W m$^{-2}$ $\mu$m$^{-1}$) of $\lambda$ Gru, deduced from the JP11-UBVRI and the 2MASS-JHK$_s$ magnitudes, and from the IRAS flux measurements at 12, 25 and 60 $\mu$m. The thin curve is the spectrum of a 4 250-K blackbody radiator with an angular diameter of 2.7 mas, given for comparison. The lengths of the short vertical bars are the values of the actual flux errors.

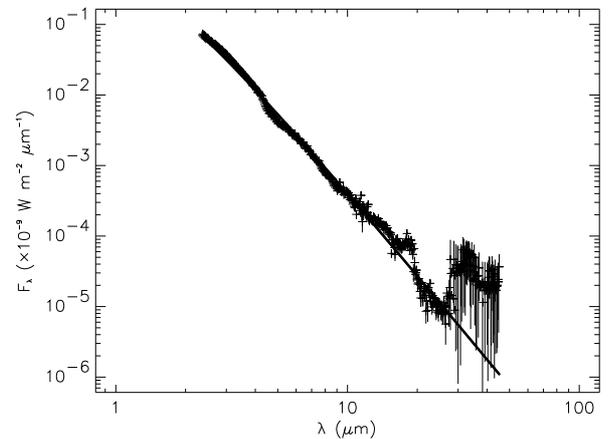

**Fig. 4.** Log-log plot of the high-resolution processed SWS01 spectrum (in nW m$^{-2}$ $\mu$m$^{-1}$) of $\lambda$ Gru from the NASA/IPAC Infrared Data Archive. The thick curve is the spectrum of a 4 250-K blackbody radiator with an angular diameter of 2.7 mas, given for comparison.

## 5. Direct and indirect approaches

To determine the stellar angular diameters, two different methods are commonly used, classified as direct and indirect by Fracassini et al. (1981). The direct method consists in linking the high angular or spectral resolution observations of some physical phenomena directly with the stellar disk geometry. Unless the instrumental response is known with an extreme accuracy, which is an extremely difficult challenge in the presence of (terrestrial) atmospheric turbulence, the accurate estimation of the calibrator angular diameters with the direct method needs very careful calibration with other unresolved or extremely well-known calibrators. In this case, the problem can be solved thanks to global calibrating strategies (Meisner, 2008; Richichi et al., 2009).

The indirect method is based on the luminosity formula $\mathcal{L} \propto \mathcal{R}^2\,T_e^4$, where $\mathcal{R}$ is the linear radius. High-fidelity SED templates (Boden, 2007) or stellar atmosphere models can be used to provide homogeneous diameter estimates. Because of the very small number of existing absolute primary standards (Cohen et al., 1992), indirect diameter estimation needs to beware of the calibration of the absolute flux, hence of the effective temperature.

The calibrator catalogues of Bordé et al. (2002, hereafter B02), Mérand et al. (2005), and van Belle et al. (2008) use this method, the first two based on the previous absolute spectral calibration works of Cohen et al. (1999), the latter based on the works of Pickles (1998). The angular diameter estimates contained in the calibrator catalogue for VLTI-MIDI MCC[9] are also inferred from the indirect method, fitting global photometric measurements by stellar atmosphere models, giving diameter uncertainties within ±5% (Verhoelst, 2005).

The compilation of all stellar diameter values published in the literature has been carried out to build the CADARS (Fracassini et al., 1981; Pasinetti Fracassini et al., 2001) and the CHARM/CHARM2 (Richichi & Percheron, 2002; Richichi et al., 2005) catalogues. Although this approach seems attractive, because it gives the impression of providing "reliable" and well-controlled diameters, a sharper analysis of the data shows that these catalogues are intrinsically heterogeneous, with a precision rarely reaching 5%.

The studies presented in the present paper follow the indirect method of estimating the angular diameters of the interferometric calibrators, comparing the results obtained with various observations: diameter from the spectral type, from the colour index, from the broadband infrared magnitude, from the Johnson photometry, and from the spectral energy distribution. We especially focus attention on determining diameter uncertainties.

## 6. Model atmospheres

Thanks to the considerable progress made in modelling the stellar atmospheres, extensive grids of synthetic fluxes and spectra are now available. To get a summary of the existing synthetic spectra, one can look, for example, at Carrasco's web page[10]. Among all the stellar atmosphere grids available, we should particularly mention: the ATLAS models[11] (Kurucz, 1979; Castelli & Kurucz, 2003), the PHOENIX stellar and planetary atmosphere code[12] (Hauschildt, 1992; Brott & Hauschildt, 2005), and the MARCS stellar atmosphere models[13] (Gustafsson et al., 1975, 2008). These codes have been compared by Kučinskas et al. (2005) for the late-type giants, and Meléndez et al. (2008) have shown the very good agreement between them. Concerning MARCS, Decin et al. (2000) has studied the influence of various stellar parameters on the synthetic spectra.

Because the MARCS code is particularly suitable for the cool stars (Gustafsson et al., 2003), we naturally opt to use it to model the atmosphere of $\lambda$ Gru. Detailed information about the models can be found on the MARCS web site. The library supplies high-resolution ($R = \lambda/\Delta\lambda = 20\,000$) energy fluxes for $0.13 \leq \lambda \leq 20\,\mu$m, for a wide grid of spherical atmospheric models, obtained with $2\,500 \leq T_e \leq 8\,000$ K (step 100 K or 250 K), surface gravities $-1.0 \leq \log g \leq 4.0$ (step 0.5), metallicities $-5.0 \leq z = $ [Fe/H] $\leq 1.0$ (with variable step from 1.0 to 0.25 dex), stellar masses $\mathcal{M}$ of 0.5, 1.0, 2.0, and 5.0 $\mathcal{M}_\odot$, and microturbulent velocity $\xi_t = 2.0$ or $5.0$ km s$^{-1}$. Figure 5 shows the high-resolution synthetic spectral radiant exitance of a typical K3III star, with $T_e = 4\,250$ K, $\log g = 2.0$, $z = 0.0$ dex, $\mathcal{M} = 1.0\,\mathcal{M}_\odot$, and $\xi_t = 1.0$ km s$^{-1}$, given by the spherical MARCS model.

[9] ster.kuleuven.ac.be/~tijl/MIDI_calibration/mcc.txt
[10] www.am.ub.es/~carrasco/models/synthetic.html
[11] kurucz.harvard.edu/
[12] www.hs.uni-hamburg.de/EN/For/ThA/phoenix/
[13] marcs.astro.uu.se/

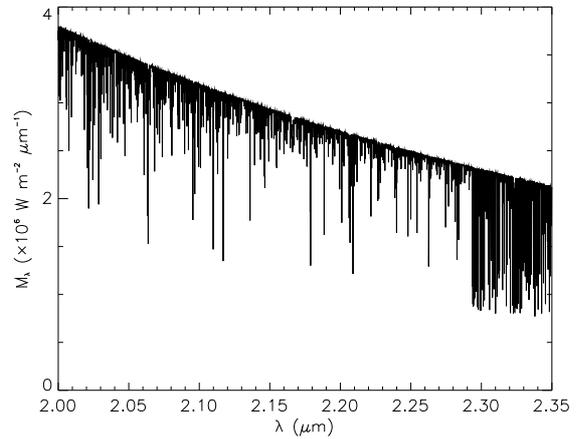

**Fig. 5.** High-resolution ($R = 20\,000$) MARCS synthetic spectral radiant exitance (in MW m$^{-2}$ $\mu$m$^{-1}$), in the K-band (2.0 to 2.35 $\mu$m), obtained with $T_e = 4\,250$ K, $\log g = 2.0$, $z = 0.0$ dex, $\mathcal{M} = 1.0\,\mathcal{M}_\odot$, and $\xi_t = 2.0$ km s$^{-1}$.

Figure 6 shows the synthetic spectral radiance, obtained with the same set of physical parameters using the TURBOSPECTRUM code (Alvarez & Plez, 1998), with a spectral step of 20 Å. In the upper left panel, the radiance spectral distribution at the disk centre is shown for $1.4 \leq \lambda \leq 2.5\,\mu$m. The upper right panel shows the radiance normalized to the centre, for various values of the distance from the star centre $r$ (expressed in photospheric radius units). The model reproduces the change from absorption (on the disk) to emission (just beyond the continuum limb) of the first overtone ro-vibrational CO band at 2.3 $\mu$m, also seen in the near-infrared solar observations (Prasad, 1998). The lower left panel shows the normalized radiance profiles for various wavelengths. The position of the inflexion point gives the wavelength-independent Rosseland to limb-darkened conversion factor $C_{\text{Ross/LD}} = \mathcal{R}(\tau_{\text{Ross}}=1)/\mathcal{R}$, where $\mathcal{R}$ is the model outermost linear diameter (Wittkowski et al., 2004). For a discussion of the different definitions of the stellar radius, one can refer to Baschek et al. (1991). The lower right panel shows first partial derivatives with respect to $r$ of the normalized radiance, against the viewing angle cosine $\mu = \sqrt{1-r^2}$. The median value 0.991 of the inflexion point is very close to the value $C_{\text{Ross/LD}} = 0.989$ predicted by the MARCS code.

For comparison purpose, we use the Planck and the Engelke (Engelke, 1992; Marengo, 2000) formulae. Representing the simplest way to model a stellar flux, the Planck function describes the exitance of a blackbody radiator with temperature $T$. Improving upon the blackbody description of the cool star infrared emission by incorporating empirical corrections for the main atmospheric effects, the Engelke function is obtained by substituting $T$ with $0.738\,T\,[1 + 79\,450/(\lambda T)]^{0.182}$ in the expression of the Planck formula ($T$ in K and $\lambda$ in $\mu$m). Because it is an analytical approximation of the 2-60 $\mu$m continuum spectrum for giants and dwarfs with $3\,500 < T_e < 6\,000$ K, the Engelke function is based on the scaling of a semi-empirical plane-parallel solar atmospheric exitance profile for various effective temperatures (Decin & Eriksson, 2007).

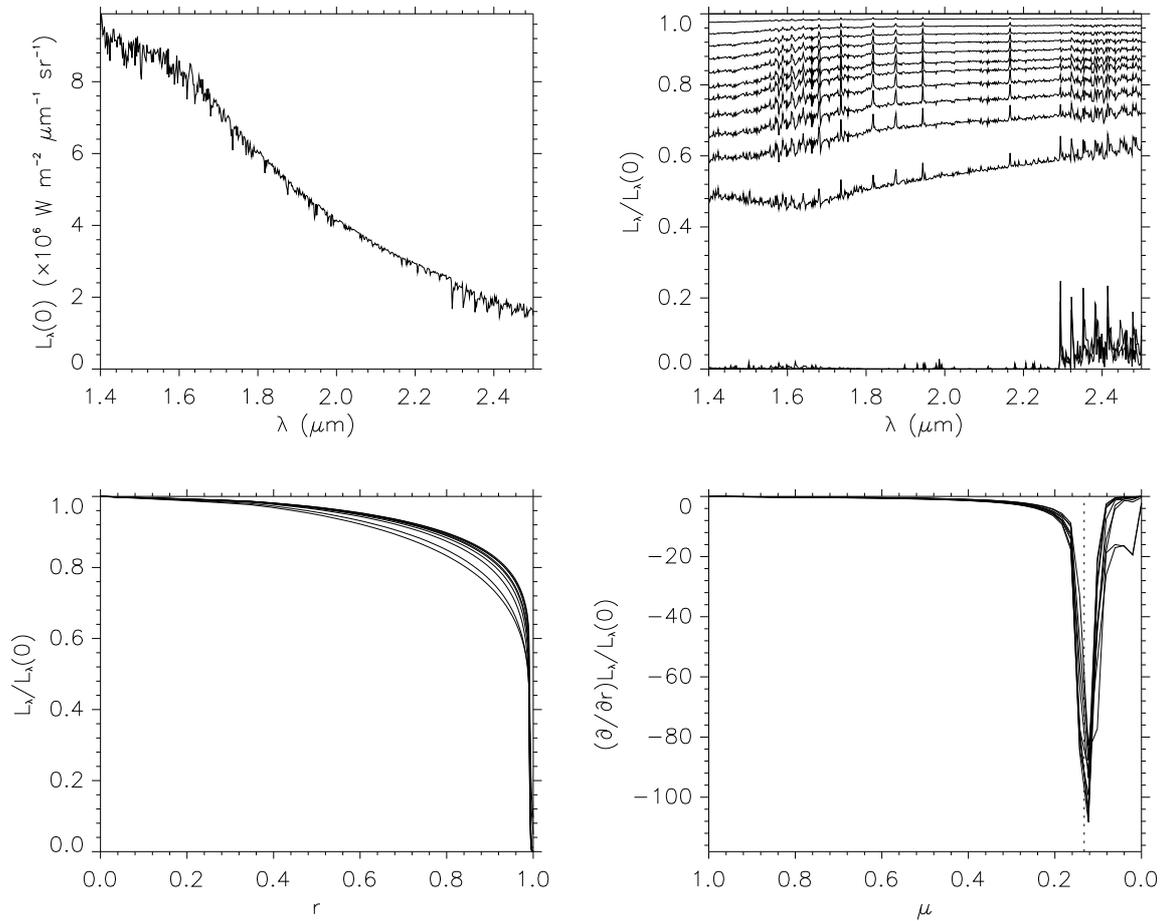

**Fig. 6.** Median-resolution ($R = 1\,000$) TURBOSPECTRUM synthetic radiance data, obtained with the same model parameters as for Fig. 5. Upper left panel: spectral distribution of the central radiance. Upper right panel: spectral distribution of the radiance normalized to the centre, for $r = 0.345$ (upper curve), 0.515, 0.631, 0.720, 0.791, 0.848, 0.883, 0.922, 0.952, 0.974, 0.990, and 0.998 (lower curve). Lower left panel: normalized radiance profiles, for the wavelengths $\lambda = 2.5$ (upper curve), 2.364, 2.226, 2.088, 1.950, 1.812, 1.674, 1.536, and $1.4\,\mu$m (lower curve). Lower right panel: partial derivatives of the normalized radiance profiles with respect to $r$, against $\mu = \sqrt{1-r^2}$. The dashed vertical line gives the median value of $C_{\text{Ross/LD}}$ (see text for details).

## 7. Diameter estimation

Among all the indirect approaches used to estimate the angular diameter, we compare now some of the most widely used methods.

### 7.1. From the spectral type

The stellar fundamental parameters mass $\mathcal{M}$, linear radius $\mathcal{R}$, and absolute luminosity $\mathcal{L}$ are directly related to the stellar atmospheric parameters effective temperature $T_e$, surface gravity $g$, according to the logarithmic formulae (Straižys & Kuriliene, 1981; Smalley, 2005):

$$\log g \approx \log \mathcal{M}/\mathcal{M}_\odot + 4\log T_e + 0.4\,M_{\text{bol}} - 12.505, \quad (8)$$
$$\log \mathcal{R}/\mathcal{R}_\odot \approx 8.471 - 2\log T_e - 0.2\,M_{\text{bol}}, \quad (9)$$
$$\Rightarrow \log g \approx \log \mathcal{M}/\mathcal{M}_\odot - 2\log \mathcal{R}/\mathcal{R}_\odot + 4.437. \quad (10)$$

This uses the solar parameter values $T_{e,\odot} = 5\,777(10)\,$K (Smalley, 2005), $\log g_\odot = 4.4374(5)$ (Gray, 2005), and $M_{\text{bol},\odot} = +4.738(1)$ deduced from the solar luminosity $\mathcal{L}_\odot = 3.8427(14)\times 10^{26}\,$W with the value $3.02 \times 10^{28}\,$W, used as the zero point of the absolute bolometric magnitude scale (Amsler et al., 2008).

To estimate the effective temperature and the luminosity from the Morgan-Keenan-Kellman (MKK) type, de Jager & Nieuwenhuijzen (1987) introduce the continuous variables $s$ (linked to the spectral class) and $b$ (linked to the luminosity) and derive mathematical expressions of $\log T_e$ and $\log \mathcal{L}/\mathcal{L}_\odot$ against $s$ and $b$, with $1\sigma$ values of 0.021 and 0.164 respectively.

For a K3III star, with $s = 5.8$ and $b = 3.0$, the two-dimensional B-spline interpolation of the tables of de Jager & Nieuwenhuijzen (1987) gives $\log T_e = 3.629(21)$ and $\log \mathcal{L}/\mathcal{L}_\odot = 1.899(164)$, hence $M_{\text{bol}} = -0.01(41)$, so that Eq. (9) gives $\log \mathcal{R}/\mathcal{R}_\odot = 1.215(92)$. To avoid the bias on the distance appearing from the inversion of the parallax (Brown et al., 1997; Luri & Arenou, 1997), we deduce the angular diameter $\phi$ from the linear radius $\mathcal{R}$ and from the parallax $\varpi$ ($\phi$ and $\varpi$ in same units), according to the relation (Allende Prieto, 2001)

$$\log \phi \approx -2.031 + \log \mathcal{R}/\mathcal{R}_\odot + \log \varpi, \quad (11)$$

based on the latest value of the angular diameter of the Sun seen at 1 pc: $\phi_{\odot,1\text{pc}} = 0.0092984(4)''$ (Amsler et al., 2008). Combining Eqs. (11) and (9) leads to a logarithmic variant of the formula giving the effective temperature $T_e$ (in K)

$$\log T_e \approx 2.746 + 0.25\log \mathcal{L}/\mathcal{L}_\odot + 0.5\log \varpi/\phi. \quad (12)$$

If the relative uncertainty on $\log \mathcal{R}/\mathcal{R}_\odot$ is higher than 20%, we follow the confidence interval transformation princi-ple (CITP), described in Appendix A, to get a rough estimate of the uncertainty on $\mathcal{R}/\mathcal{R}_\odot$. Using the value $\log \mathcal{R}/\mathcal{R}_\odot = 1.215(92)$ derived above and the parallax $13.20(78)$ mas, we obtain from Eq. 11 $\phi = 2.01 \binom{57}{45}$ mas.

The angular diameter estimate given by this method clearly underestimates the B02 value (2.71 mas). An incorrect parallax value given by Hipparcos cannot be suspected, considering the relative proximity of $\lambda$ Gru, located at 76(4) pc. Slight errors in determining the luminosity class could be a more likely cause of bias in diameter estimation. For $\lambda$ Gru, we find that a luminosity $\log \mathcal{L}/\mathcal{L}_\odot = 2.253(164)$ would be more adequate than $\log \mathcal{L}/\mathcal{L}_\odot = 1.899(164)$, giving $\phi = 3.00\binom{86}{67}$ mas.

### 7.2. From the colour index

Because the accurate stellar classification is a very difficult challenge leading to potential misclassifications, other parameters must be used to investigate the relation between the stellar temperature and the angular size. Being relatively independent of stellar gravities and abundances, the NIR colours are very good temperature indicators (Bell & Gustafsson, 1989). For cool stars, the $V - K$ colour index is also known to be the most appropriate parameter for representing the apparent bolometric flux, almost independently of their luminosity class (Johnson, 1966; di Benedetto, 1993), The empirical derivation of the angular sizes from the colour indices have been studied by many authors (di Benedetto, 1998; van Belle, 1999; Groenewegen, 2004), leading to different relations. For our study, we use the following relations proposed by van Belle et al. (1999), particularly suitable for late-type giants and supergiants

$$T_e \approx 3\,030 + 4\,750 \times 10^{-0.187(V-K)}, \qquad (13)$$
$$\log \mathcal{R}/\mathcal{R}_\odot \approx 0.245 + 2.36 \log(V - K). \qquad (14)$$

The average standard deviations are: 250 K on $T_e$, and 30% on $\mathcal{R}/\mathcal{R}_\odot$. One of the major difficulties with this method is the correction of the colour index for the interstellar absorption. Appendix B briefly describes the de-reddening process used in our study. Table 2 gives the results of the correction for the interstellar extinction in the Johnson and in the 2MASS bands. The UBVRI magnitudes come from the JP11 Catalogue (Morel & Magnenat, 1978). Because data precision may vary significantly (Nagy & Hill, 1980), a conservative value of 0.05 has been arbitrarily chosen as the uncertainty on each magnitude. The $JHK_s$ magnitudes and uncertainties are taken from the 2MASS Catalogue (Cutri et al., 2003).

Using the corrected (intrinsic) $V - K$ colour index 3.16(29) deduced from Table 2, we can infer from Eqs. (13) and (14) that the effective temperature of $\lambda$ Gru is 4 247(250) K, and the linear radius is 26.6(80) $\mathcal{R}_\odot$. Since the linear radius relative uncertainty is 30%, we estimate the angular diameter uncertainty range according to the CITP.

As a result, with the parallax 13.20(78) mas, the $V - K$ angular diameter is 3.3(10) mas, slightly greater than the B02 value 2.71 mas. A $V - K$ value 2.92 would give an angular diameter estimate closer to the B02. Unfortunately, the high level of final uncertainties prevent knowing the most likely source of bias: errors on the input magnitudes, de-reddening process, or diameter estimation method itself.

The angular diameter estimation given by the JMMC-SearchCal tool is, for bright objects (Bonneau et al., 2006), based on the study undertaken by Delfosse (2004), where a least-square polynomial fit of the distance-independent diameter esti-mator $\psi_V = \sum_k a_k (CI)^k$ against each deredenned colour index $CI$ is achieved. Introducing the distance modulus $m_V - M_V = -5 \log \varpi - 5$ (with $\varpi$ in arcsec) in Eq. (11), where $m_V$ and $M_V$ are the apparent and the absolute stellar magnitudes in V, one can define $\psi_V$ by

$$\log \psi_V = \log \mathcal{R}/\mathcal{R}_\odot + 0.2 M_V - 1 \approx \log \phi + 0.2 m_V - 0.969, \quad (15)$$

for $\phi$ in mas. Among the empirical relations between $\psi_V$ and each colour index, the highest accuracies on the angular diameter $\phi$ given by Eq. (15) are obtained with the three colour indices $B - V$ ($\Delta\phi/\phi = 8\%$, for $-0.4 \leq CI \leq 1.3$), $V - R$ (10%, for $-0.25 \leq CI \leq 2.8$), and $V - K$ (7%, for $-1.1 \leq CI \leq 7.0$). Unlike the classical methods of angular diameter estimation from the colour index, as the method of van Belle et al. (1999), which needs a parallax estimate in complement to magnitude measurements in 2 bands, Bonneau's method needs only photometric data, more precisely the apparent $V$ magnitude and the colour indices. With $B - V = 1.36(25)$, $V - R = 0.99(19)$, and $V - K = 3.16(21)$, the corresponding angular diameter estimates of $\lambda$ Gru are $\phi_{B-V} = 3.32(27)$ mas, $\phi_{V-R} = 2.97(30)$ mas, and $\phi_{V-K} = 3.01(21)$ mas. Although this method gives coherent diameter estimates within ±11%, which confirms that *SearchCal* considers $\lambda$ Gru as a suitable calibrator for interferometry, it also overestimates the B02 value 2.71 mas, especially using the $B - V$ colour index. For a K3 giant, the fiducial Johnson colours taken from spectral type-luminosity class-colour relations given by Bonneau et al. (2006), would be: $B - V = 1.27$, $V - R = 0.98$, and $V - K = 3.01$. With these colour indices, the angular diameter estimates would be: $\phi_{B-V} = 2.89$, $\phi_{V-R} = 2.93$, and $\phi_{V-K} = 2.78$, close to the B02 value. At least 3 causes of bias may be suspected:

1. Decreasing the $B$ corrected magnitude from 5.79 to 5.70 would be sufficient to lower the $B - V$ angular diameter estimate to 2.94 mas, so that the diameter estimates in $B - V$, $V - R$, and $V - K$, would stay within ±2%. Thus, a slight overestimate of the $B$ magnitude of $\lambda$ Gru in the JP11 catalogue may be suspected.
2. On the other hand, tests of the de-reddening procedure show that, even if we artificially increase the visual extinction from 0.04 to an unrealistically high value of 2.0 mag, the angular diameter derived from $B - V$ would decrease from 3.32 to barely 3.30 mas, while the $V - R$ and the $V - K$ diameters would get closer to the B02 value, respectively from 2.97 to 2.65 mas, and from to 3.01 to 2.84 mas.

3. Since the intrinsic $B - V$ colour index of $\lambda$ Gru (1.36) is slightly larger than the upper limit (1.30) of the validity domain of the polynomial fit, it is finally not surprising that the method gives an incorrect diameter estimate from $B - V$ in this case.

**Table 2.** Broadband photometry of $\lambda$ Gru and reddening.

| $\lambda_0[W_0]^a$ | m$^b$ | R$^c$ | m$_{cor}{}^d$ | F$_0{}^e$ | F$_{band}{}^f$ |
|---|---|---|---|---|---|
| 0.35 [0.07] | 7.49(5) | 4.87 | 7.44(24) | 41.75(84) | 0.05(1) |
| 0.44 [0.09] | 5.83(5) | 4.02 | 5.79(20) | 63.20(13) | 0.31(6) |
| 0.55 [0.09] | 4.46(5) | 3.01 | 4.43(15) | 36.31(73) | 0.62(9) |
| 0.69 [0.21] | 3.46(5) | 2.16 | 3.44(11) | 21.77(44) | 0.92(10) |
| 0.88 [0.23] | 2.68(5) | 1.46 | 2.67(9) | 11.26(23) | 0.97(8) |
| 1.24 [0.16] | 2.10(26) | 0.81 | 2.09(26) | 3.13(5) | 0.47(11) |
| 1.65 [0.25] | 1.44(25) | 0.52 | 1.43(25) | 1.13(2) | 0.31(7) |
| 2.17 [0.26] | 1.27(25) | 0.35 | 1.27(25) | 0.43(8) | 0.14(3) |

$^a$ filter mean wavelength and equivalent width (both in $\mu$m)
$^b$ measured in-band magnitude
$^c$ ratio of total to selective extinction
$^d$ de-reddened in-band magnitude, adopting $A_V = 0.03(15)$
$^e$ zero-magnitude flux (in nW m$^{-2}$ $\mu$m$^{-1}$)
$^f$ in-band mean absolute flux (in nW m$^{-2}$ $\mu$m$^{-1}$)

### 7.3. From the in-band magnitude

The two methods for estimating the stellar diameter presented above are based on statistical relations and do not use any photospheric model. On the contrary, the methods presented in the following sections explicitly need photospheric models. As first shown by Blackwell & Shallis (1977), the photometric angular diameter in a spectral band can be estimated thanks to the relation $\phi_{\text{band}} = 2\sqrt{F_{\text{band}}/M_{\text{band}}}$, where $F_{\text{band}}$ and $M_{\text{band}}$ are the received and emergent mean fluxes in the considered band (both in W m$^{-2}$ $\mu$m$^{-1}$). The angular diameters derived with this method, known as the infra-red flux method (IRFM), are generally accurate to between 2 and 3% (Blackwell et al., 1990), depending not only on the fidelity of the atmospheric models used in the calibration, but also on the uncertainty in the absolute flux determination.

The last column of Table 2 lists the received absolute fluxes deduced from the measured de-reddened in-band magnitudes $m_{\text{cor}} = -2.5 \log F_{\text{band}}/F_0$, where $F_0$ is the zero-magnitude flux taken from Bessell et al. (1998) for UBVRI, with 2% uncertainties (Colina et al., 1996), and from Cohen et al. (2003) for JHK$_s$. For in-band corrected magnitudes with relative uncertainties exceeding 20%, absolute flux uncertainties are computed according to the CITP (see Eq. A.3).

If $t(\lambda)$ is the transmission profile of the considered filter, normalized to 1.0 at its maximum, one can define the in-band effective wavelength and width as (Fiorucci & Munari, 2003)

$$\lambda_e = \frac{\int \lambda \, t(\lambda) \, M(\lambda) \, d\lambda}{\int t(\lambda) \, M(\lambda) \, d\lambda}, \quad (16)$$

$$\int_{\lambda_e - W_e/2}^{\lambda_e + W_e/2} M(\lambda) \, d\lambda = \int t(\lambda) \, M(\lambda) \, d\lambda, \quad (17)$$

so that the band emergent mean flux $M_{\text{band}}$ can be written as

$$M_{\text{band}} = \frac{\int t(\lambda) \, M(\lambda) \, d\lambda}{\int t(\lambda) \, d\lambda} = \frac{\int_{\lambda_e - W_e/2}^{\lambda_e + W_e/2} M(\lambda) \, d\lambda}{W_0}, \quad (18)$$

where $W_0 = \int t(\lambda) \, d\lambda$ is the equivalent width of the band transmission profile. The $W_0$ values are enclosed in square brackets in Table 2. Introducing

$$\tilde{M}_{\text{band}} = \frac{\int_{\lambda_e - W_e/2}^{\lambda_e + W_e/2} M(\lambda) \, d\lambda}{W_e}, \quad (19)$$

the in-band angular diameter is given by $\phi_{\text{band}} = 2\sqrt{\gamma_{\text{band}} F_{\text{band}}/\tilde{M}_{\text{band}}}$, where $\gamma_{\text{band}} = W_0/W_e$. Depending on the model used, the effective band parameters $\lambda_e$ and $W_e$, presented in Table 3 for a K2 spectrum (4 380 K) and for a 4 250 K blackbody spectrum, are extracted from the Asiago Database on Photometric Systems.[14] Table 4 gives the in-band angular diameters using the Planck, the Engelke, and the MARCS synthetic spectra with the same temperature of 4 250 K, integrated in the 2MASS J, H, and K$_s$ spectral bands. When the absolute flux uncertainties are greater than 20%, we compute the angular diameter uncertainties according to the CITP.

**Table 3.** Effective wavelength and bandwidth (in square brackets) in each band (both in $\mu$m), for a K2 spectral type spectrum and a 4 250 K blackbody spectrum.

| Filter name | K2 spectrum | 4 250 K Planck |
|---|---|---|
| Johnson-U | 0.361 [0.059] | 0.359 [0.061] |
| Johnson-B | 0.457 [0.079] | 0.451 [0.090] |
| Johnson-V | 0.557 [0.085] | 0.556 [0.086] |
| Johnson-R | 0.691 [0.205] | 0.693 [0.204] |
| Johnson-I | 0.870 [0.230] | 0.869 [0.229] |
| 2MASS-J | 1.240 [0.160] | 1.239 [0.166] |
| 2MASS-H | 1.640 [0.250] | 1.646 [0.248] |
| 2MASS-K$_s$ | 2.150 [0.260] | 2.154 [0.254] |

**Table 4.** Photometric angular diameters (in mas) obtained with various synthetic spectra (with $T = 4\,250$ K) in the 2MASS near-infrared bands.

| Model | $\phi_J$ | $\phi_H$ | $\phi_{K_s}$ |
|---|---|---|---|
| Planck | 3.00$\binom{39}{44}$ | 3.35$\binom{40}{46}$ | 3.22$\binom{38}{43}$ |
| Engelke | 2.24$\binom{29}{33}$ | 2.80$\binom{34}{39}$ | 2.92$\binom{35}{39}$ |
| MARCS | 2.87$\binom{37}{42}$ | 2.78$\binom{34}{38}$ | 2.80$\binom{33}{38}$ |

Given for comparison, the overestimated angular diameters obtained with the Planck spectrum confirm that the stellar photospheres may deviate noticeably from simple blackbodies. Similarly, the underestimated J-band angular diameter derived from the Engelke spectrum confirms that the Engelke analytic approximation is valid for wavelengths longer than 2 $\mu$m. Finally, the MARCS synthetic spectrum with $T_e = 4\,250$ K, $\log g = 2.0$, $z = 0.0$ dex, $\mathcal{M} = 1.0 \mathcal{M}_\odot$, and $\xi_t = 2.0$ km s$^{-1}$ yields angular diameters in $J$, $H$, and $K_s$, which are close to the B02 value of 2.71 mas, and with less dispersion.

### 7.4. From the broadband photometry

The IRFM method, described in Sect. 7.3, gives different angular-diameter estimates for each spectral band in which the model spectrum is integrated. To get a unique estimate of the angular diameter, taking the global broadband photometry into account (as shown for example in Fig. 3), the use of fitting techniques is necessary. The most widely used is based on $\chi^2$ minimization.

If $\sigma_i$ is the measurement error of the mean flux $F_i$, received in spectral band $i$, and $M_i$ the emergent mean flux in the same band, the best-fit angular diameter corresponds to the minimum of the one-parameter $\chi^2(\phi)$ nonlinear function defined by

$$\chi^2(\phi) = \frac{1}{N-1} \sum_{i=1}^{N} \frac{1}{\sigma_i^2} \left( F_i - \frac{\phi^2}{4} M_i \right)^2, \quad (20)$$

where $N$ is the total number of spectral bands used to build the global photometry. To find the minimum value of the $\chi^2$ function, we use the gradient-expansion algorithm, which combines the features of the gradient search with the method of linearizing the fitting function (Bevington & Robinson, 1992), very similar to the classical Levenberg-Marquardt algorithm (Levenberg, 1944; Marquardt, 1963). Figure 7 shows an example of $\chi^2$ against the angular diameter obtained when fitting the MARCS model on the ISO SWS data, as described in Sect. 7.5, using the gradient-expansion algorithm. First used by Cohen et al.

---
[14] ulisse.pd.astro.it/Astro/ADPS/

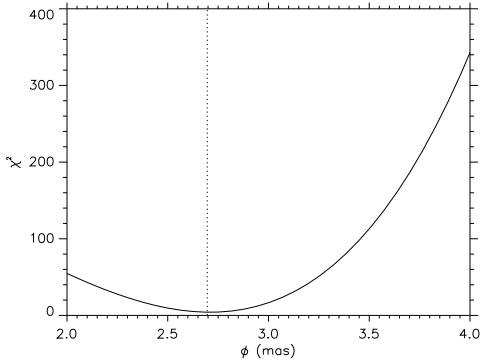

**Fig. 7.** Plot of $\chi^2$ against the angular diameter $\phi$ (in mas) obtained by fitting the appropriate MARCS model on the SWS data (see Sect. 7.5), using the gradient-expansion algorithm. The vertical dotted line gives the position of the best-fit parameter.

**Table 5.** Angular diameters obtained by fitting various models (with $T = 4\,250$ K) to visible and/or NIR photometric data.

| Used bands | Model | $\phi_{\text{best}}$[a] | $\sigma_{\text{fit}}$[b] | $\chi^2_{\text{min}}$[c] | F2$_{\text{fin}}$[d] | $\phi$-range[e] |
|---|---|---|---|---|---|---|
| $B...K_s$ | Planck | 2.91 | 0.10 | 2.30 | 1.85 | 2.45-3.19 |
| $U...K_s$ | MARCS | 2.83 | 0.10 | 2.18 | 1.85 | 2.37-3.27 |
| $JHK_s$ | Planck | 3.18 | 0.24 | 0.18 | -0.98 | – |
| $JHK_s$ | Engelke | 2.54 | 0.20 | 1.26 | 0.58 | – |
| $JHK_s$ | MARCS | 2.80 | 0.21 | 0.02 | -1.80 | – |

[a] best-fit diameter (in mas)
[b] formal fit error (in mas)
[c] minimum $\chi^2$ value
[d] final goodness-of-fit parameter
[e] 68% confidence interval (in mas)

(1992) for the absolute calibration of broad- and narrow-band infrared photometry, based upon the Kurucz stellar models of Vega and Sirius, this method has led to the construction of a self-consistent, all-sky network of over 430 infrared radiometric calibrators (Cohen et al., 1999), upon which the further works of B02 and Mérand et al. (2005) are based. Estimating the stellar angular diameters through photometric modelling is also used in the MSC-getCal Interferometric Observation Planning Tool Suite, which relies on the Planck blackbody SED, parameterized by its effective temperature and bolometric flux (see the *fbol* routine in the reference manual available online[15]).

Table 5 gives the results of the fitting process for the visible and/or NIR photometry (given in Table 2), with the Planck, the Engelke (suitable for infrared wavelengths only), and the MARCS models. Since the blackbody model ignores line-blanketing effects in the near-UV (Allende Prieto & Lambert, 2000), as seen in Fig. 3, the Johnson-$U$ flux is not considered when fitting the Planck model. The best-fit angular diameters correspond to the minimum values of the $\chi^2$ function. To compare the $\chi^2_{\text{min}}$ values obtained for data samples with different sizes, it is convenient to use the $F2$ goodness-of-fit parameter defined as (Kovalevsky & Seidelmann, 2004)

$$F2 = \left(\frac{9\nu}{2}\right)^{1/2}\left[\left(\frac{\chi^2}{\nu}\right)^{1/3} + \frac{2}{9\nu} - 1\right], \quad (21)$$

where $\nu$ is the number of degrees of freedom, equal to $N - 1$ in our case (1 parameter). When $\nu$ gets larger than 20, $F2$ tends to be normally distributed with zero mean and unit standard deviation. Bad fits correspond to $F2$ values higher than 3 (especially after removing outliers), while abnormally good fits correspond to high negative values (Jancart et al., 2005). To identify the extreme outliers, we use the upper and lower outer fences defined by $Q1 \pm 3IQR$, where $IQR = Q3 - Q1$ is the interquartile range, and $Q1$ and $Q3$ are the first and third quartiles, respectively (Zhang et al., 2004).

[15] nexsciweb.ipac.caltech.edu/gcWeb/doc/getCal/gcManual.html

As underlined by Press et al. (2007), although $\chi^2$ minimization is a useful way of estimating the parameters, the formal covariance matrix of the output parameters has a clear quantitative interpretation only if the measurement uncertainties are normally distributed. To derive robust estimates of the model parameter uncertainties, we used the confidence limits of the fitted parameters with the bootstrap method (Efron, 1979, 1982). Rather than resampling the individual observations with replacement, we use the method of residual resampling, more relevant for regression, as described in Appendix C.

In Table 5, the 68% confidence interval limits of the best-fit angular diameter are determined by bootstrap, with 1 000 resampling loops, only for data sets containing more than 5 photometric bands. Although the angular diameters listed in Table 5 (obtained by fitting model fluxes on photometric measurements) are very close to those obtained from the IRFM method, we did not consider the former results as very robust, considering the small number of photometric bands used for the fits.

### 7.5. From the spectrophotometry

Fitting atmospheric models on sparse photometric data may result in large uncertainties on the angular diameter. To decrease them significantly, larger data sets are needed. The observational data for this section consist in spectro-photometric measurements obtained with ISO-SWS (de Graauw et al., 1996). The $\lambda$ Gru spectrum shown in Fig. 4, extracted from the NASA/IPAC Infrared Science Archive, was obtained in the SWS01 observing mode (low-resolution full grating scan, on-target time = 1140 s), which covers the entire 2.4-45.4 $\mu$m SWS spectral range, with a variable spectral resolution (Table 6). The SWS AOT-1 spectra, subdivided into wavelength segments (Leech et al., 2003), have been processed and renormalized with the post-pipeline algorithm referred as the *swsmake* code (Sloan et al., 2003). The spectral characteristics of the SWS AOT-1 bands and their $1\sigma$ photometric accuracies given in Table 6 were deduced from Leech et al. (2003) and Lorente (1998). Since the spectrum of $\lambda$ Gru is very noisy at wavelengths longer than 27.5 $\mu$m, as shown in Fig. 4, probably because of calibration problems, we did not use the bands 3E and 4 for fitting the models on the ISO-SWS data.

Table 7 gives the results of the fitting process of the SWS 2.38-27.5 $\mu$m spectrum with the Planck and Engelke models (both with a temperature of 4 250 K) and with the K3III MARCS model, presented in Sect. 6. The confidence interval limits were estimated using the bootstrap resampling ($N_{\text{boot}} = 1\,000$) for the 68% confidence level. The agreement between the angular diameter obtained by spectrophotometry fitting with the MARCS model and the B02 value, obtained by fitting with the Kurucz model, reflects the excellent agreement between the two models (Meléndez et al., 2008). Deriving the angular diameters from the fit of the Engelke function on the SWS spectra (extended to 45.2 $\mu$m), Heras et al. (2002) gives an overestimated value of the angular diameter (2.82(21) mas, as compared to 2.71(3) from B02). Figure 8 shows a typical example of the histogram of the residual-bootstrap estimates of the best-fit angular diameter, ob-tained with the MARCS model on the SWS AOT-1 data, where one can see that the resulting distribution of angular diameters is notably asymmetric.

**Table 6.** Spectral characteristics of the SWS AOT-1 bands and their photometric accuracy.

| Band | $\lambda$-range[a] | R-range[b] | W-range[c] | Accuracy[d] |
|---|---|---|---|---|
| 1A | 2.38-2.60 | 711-802 | 33-32 | 4 |
| 1B | 2.60-3.02 | 559-665 | 47-45 | 4 |
| 1D | 3.02-3.52 | 665-817 | 45-43 | 4 |
| 1E | 3.52-4.08 | 490-585 | 72-70 | 4 |
| 2A | 4.08-5.30 | 585-809 | 70-66 | 7 |
| 2B | 5.30-7.00 | 353-475 | 150-147 | 7 |
| 2C | 7.00-12.0 | 475-931 | 147-129 | 7 |
| 3A | 12.0-16.5 | 475-669 | 253-247 | 12 |
| 3C | 16.5-19.5 | 669-904 | 247-216 | 10 |
| 3D | 19.5-27.5 | 372-483 | 524-570 | 19 |
| 3E | 27.5-29.0 | 494 | 559-586 | 17 |
| 4 | 29.0-45.2 | 388-619 | 748-730 | 22 |

[a] wavelength range (in $\mu$m)
[b] spectral resolution range
[c] bandwidth range (in Å)
[d] $1\sigma$ photometric accuracy (in %)

**Table 7.** Angular diameters obtained by fitting various models (with $T = 4250$ K) to the 2.38-27.5 $\mu$m SWS spectrum.

| Model | $\phi_{best}$ | $\sigma_{fit}$ | $\chi^2_{min}$ | F2$_{fin}$ | $\phi$-range |
|---|---|---|---|---|---|
| Planck | 2.70 | 0.004 | 6.5 | 38 | 2.64-2.78 |
| Engelke | 2.72 | 0.004 | 5.4 | 32 | 2.69-2.76 |
| MARCS | 2.70 | 0.004 | 4.2 | 26 | 2.66-2.79 |

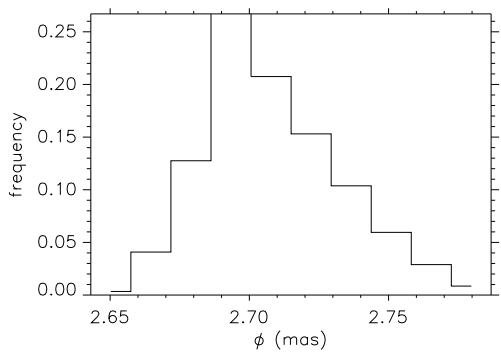

**Fig. 8.** Histogram of the angular diameters estimated by residual bootstrapping when fitting the MARCS model on the SWS spectrum.

# 8. Discussion

## 8.1. Diameter uncertainty

Figure 9 summarizes the angular diameter estimates of the test-case calibrator $\lambda$ Gru, obtained in the present study using various data types and methods. The B02 value of 2.71 mas is given for comparison. The last method (estimating the angular diameter from a fit of the SWS spectrum with a MARCS model gives, as expected, the most reliable estimate of the angular diameter, very close to the B02 estimate, with an uncertainty of 2.7%. It is also noticeable that the weighted average of the 23 angular diameter estimates is 2.73 mas.

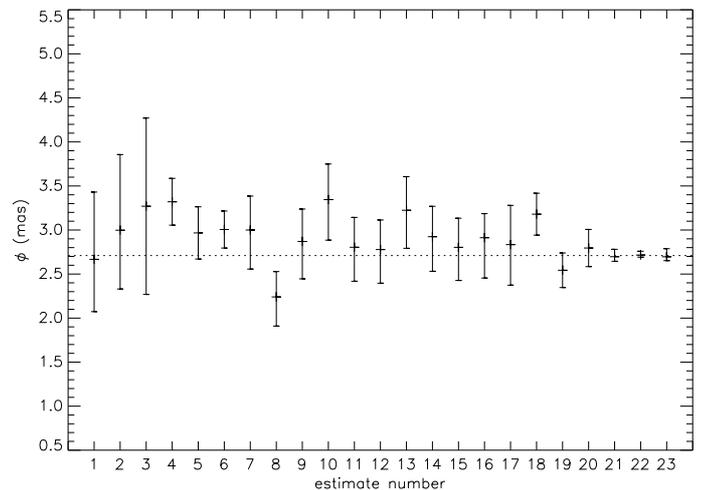

**Fig. 9.** Estimated sizes of $\lambda$ Gru with each method. The horizontal dotted line is the B02 value (2.71 mas). Estimates #1 and 2 are from the K3III spectral type using 20 and 40 terms of the polynomial expansion of de Jager & Nieuwenhuijzen (1987), #3 from $V - K$ with van Belle, #4 to 6 from $B - V$, $V - R$ and $V - K$ with Bonneau, #7 to 9 from the $J$ magnitude, #10 to 12 from $H$, and #13 to 15 from $K_s$, using the IRFM with the Planck, Engelke, or MARCS models, respectively. #16 and 17 are from the broad-band photometry with the Planck or MARCS models, #18 to 20 from the NIR photometry with the Planck, Engelke, or MARCS models; similar for #21 to 23 but from the SWS spectrum.

We must underline that the uncertainty of the limb-darkened angular diameter given by B02 is deduced from the formal standard error associated with the best-fit value of the multiplicative factor, scaling the appropriate Kurucz model on the infrared fluxes (Cohen et al., 1996). Called biases by Cohen et al. (1995), scale factor uncertainties rarely reach 1% with their method, independent of the spectral type and luminosity class. If we use, in the same manner, the formal fit errors as uncertainty estimators, the diameter uncertainty only amounts to 0.15%. Since we consider this extremely low value as unrealistic, we prefer to estimate the angular-diameter uncertainty from the statistically-significant confidence intervals, given by the residual bootstrapping (Appendix C). This amounts to 2.4% with the fit of the MARCS model.

## 8.2. Fundamental stellar parameters

From this angular-diameter accurate estimate and the parallax, we can infer a set of fundamental stellar parameters for $\lambda$ Gru, presented in Table 8, using the following procedure.

1. Calculate the linear radius $\mathcal{R}$ from $\phi$ and $\varpi$, according to Eq. (11). For $\phi = 2.70$ mas and $\varpi = 13.20$ mas, we find $\mathcal{R} = 22.0 \mathcal{R}_\odot$.
2. Fix the value of the spectral class variable $s$ from the spectral type. For a K3 star, $s = 5.80$.
3. Find the value of the luminosity class variable $b$, which gives the same angular diameter value. One can easily demonstrate from Eq. (12) that $b$ is solution of the equation

$$2t(s,b) - 0.5l(s,b) \approx 5.492 + \log \varpi - \log \phi, \qquad (22)$$

where $t(s, b)$ and $l(s, b)$ are the two-dimensional B-spline interpolation functions of the tables $\log T_e$ and $\log \mathcal{L}/\mathcal{L}_\odot$, published by de Jager & Nieuwenhuijzen (1987). For $\phi = 2.70$ mas and $\varpi = 13.20$ mas, we find $b = 2.76$.

**Table 8.** Fundamental parameter estimates (with uncertainties) of $\lambda$ Gru reevaluated from our study.

| Parameter | Estimate |
|---|---|
| $\varpi$ (mas) | 13.20(78) |
| $\phi$ (mas) | $2.70\binom{9}{4}$ |
| $\mathcal{R}/\mathcal{R}_\odot$ | 22.3(32) |
| $T_e$ | 4 269(206) |
| $\mathcal{L}/\mathcal{L}_\odot$ | 155(53) |
| $M_{bol}$ | -0.66(41) |
| $\log g$ | 2.13(40) |
| $\mathcal{M}/\mathcal{M}_\odot$ | $2.4\binom{61}{17}$ |

4. From $s = 5.80$ and $b = 2.76$, deduce the interpolated effective temperature $T_e$ and the absolute luminosity $\mathcal{L}$, and calculate the bolometric magnitude $M_{bol} = 4.138 - 2.5 \log \mathcal{L}/\mathcal{L}_\odot$.
5. Interpolate the surface gravity $\log g$ in the corresponding table of Straižys & Kuriliene (1981) for the same couple of values $(s, b)$.
6. Using Eq. (10), deduce the stellar mass $\mathcal{M}$ from $\mathcal{R}$ and $\log g$.

Using this method, one can see from Table 8 that the bolometric magnitude and especially the stellar mass are determined with very low accuracies. The fundamental parameter accuracies are computed using the $1\sigma$ accuracies given by de Jager & Nieuwenhuijzen (1987): 0.021 for $\log T_e$, 0.164 for $\log \mathcal{L}/\mathcal{L}_\odot$, and 0.4 for $\log g$. The uncertainties on the fundamental parameters deduced by our study validate a posteriori the choice of the input parameter values for the MARCS model used to fit the flux measurements: $T_e = 4250$ K, $\log g = 2.0$, and $\mathcal{M} = 1.0\,\mathcal{M}_\odot$.

### 8.3. Model parameters

One critical point of our method is the preliminary choice of a single set of photospheric model parameters, used to infer the angular diameter. To determine them accurately, we refer the reader to the papers by Decin et al. (2000) and Decin et al. (2004). In our study, we use the MARCS model with the fiducial parameter set of a K3III star, i.e. $T_e = 4250$ K, $\log g = 2.0$, $z = 0.0$ dex, $\mathcal{M} = 1.0\,\mathcal{M}_\odot$, and $\xi_t = 2.0$ km s$^{-1}$. The very good agreement between the angular diameter estimates deduced from the fit of the Planck, the Engelke and the MARCS models on the ISO-SWS 2.38-27.5 $\mu$m spectrum of $\lambda$ Gru, as it is shown in Table 7, justifies the choice of these model parameters.

## 9. Conclusion

In our paper, we have compared different methods for angular-diameter estimation of the interferometric calibrators. The spectral-type angular diameters only need distances as extra input. The colour-index diameters need a good interstellar correction. The photometric and the spectrophotometric diameters need explicit synthetic spectra. As expected, the results are highly dependent on the number and quality of the input data.

As a test case, we used the giant cool star $\lambda$ Gru that we observed to calibrate the VLTI-AMBER low-resolution $JHK$ observations[16] of the scientific target $\pi^1$ Gru (Sacuto et al., 2008), which will be the subject of our forthcoming paper.

Each diameter estimate is compared to the B02 value (2.71 mas) found in the Catalogue of Calibrator Stars for

---
[16] ESO programme ID 080.D-0076A (AMBER GTO)

Interferometry. The most reliable estimate of the angular diameter we find is 2.70 mas, with a 68% confidence interval 2.65-2.81 mas, obtained by fitting the ISO/SWS spectrum (2.38-27.5 $\mu$m) with a MARCS atmospheric model, characterized by $T_e = 4250$ K, $\log g = 2.0$, $z = 0.0$ dex, $\mathcal{M} = 1.0\,\mathcal{M}_\odot$, and $\xi_t = 2.0$ km s$^{-1}$. One original contribution of our study is the estimation of the statistically-significant uncertainties by means of the unbiased confidence intervals, determined by residual bootstrapping.

All numerical results and graphical outputs presented in the paper were obtained using the routines developed under PV-WAVE®, which compose the modular software suite SPIDAST[17], created to calibrate and interpret spectroscopic and interferometric measurements, particularly those obtained with VLTI-AMBER (Cruzalèbes et al., 2008). The main functionalities of the SPIDAST code, intended to be available to the community, are

1. Estimate the calibrator angular diameter by any of the methods described in this paper;
2. Create calibrator synthetic measurements, for the instrumental configuration (spectral fluxes, visibilities, and closure phases);
3. Estimate the instrumental response from the observational and the synthetic measurements of the calibrator;
4. Calibrate the observational measurements of the scientific target with the instrumental response;
5. Determine the science parameters by fitting the chromatic analytic model on the true science measurements, with the confidence intervals given by residual bootstrapping.

*Acknowledgements.* P.C. thanks A. Spang, Y. Rabbia, O. Chesneau, and A. Mérand for helpful discussions. A.J. is grateful to B. Plez and T. Masseron for their ongoing support with the MARCS code. S.S. acknowledges funding by the Austrian Science Fund FWF under the project P19503-N13. We also thank the anonymous referee whose comments helped us to improve the clarity of this paper. This research has made use of the Jean-Marie Mariotti Centre SearchCal service, co-developed by FIZEAU and LAOG, of the CDS Astronomical Databases SIMBAD and VIZIER, and of the NASA Astrophysics Data System Abstract Service.

---
[17] SPectro-Interferometric Data Analysis Software Tool

# Appendix A: Computing the uncertainties

As defined in the *Guide to the Expression of Uncertainty in Measurement* (JCGM/WG 1, 2008), the uncertainty $\sigma_X$, associated to the "best" estimate $\tilde{X}$ of a given random variable $x$, usually given by the sample average, characterizes the dispersion of $x$ about $\tilde{X}$. When it is associated to the level of confidence $1 - \alpha$, it can be interpreted as defining the interval around $\tilde{X}$, which encompasses $100(1-\alpha)\%$ of the estimates X of $x$. By analogy with the $1\sigma$ dispersion in the normal case, one can define the standard uncertainty of $\tilde{X}$ by the interval that encompasses 68.3% of the distribution of $x$ around $\tilde{X}$.

If $\sigma_{X\pm}$ are the right and left deviations of $x$ varying in the $100(1-\alpha)\%$ confidence interval (CI) about $\tilde{X}$, they can be defined as $\sigma_{X+} = X_{up} - \tilde{X}$ and $\sigma_{X-} = \tilde{X} - X_{low}$, where $X_{up}$ and $X_{low}$ are the upper and lower bounds of the CI, respectively given by the $100(1 - \alpha/2)\%$ and $100(\alpha/2)\%$ quantiles of the distribution of $x$.

To propagate the uncertainties with a monotonic transformation function $f$ of the input variable $x$ into the output variable $y = f(x)$, one can follow the confidence interval transformation principle (CITP) (Smithson, 2002; Kelley, 2007)

$$\text{Prob}\left(Y_{low} \leq y \leq Y_{up}\right) = \text{Prob}\left(X_{low} \leq x \leq X_{up}\right) = 1 - \alpha, \quad (A.1)$$

where $Y_{up}$ and $Y_{low}$ are the upper and lower bounds of the $100(1-\alpha)\%$ CI about the output best estimate $\tilde{Y} = f\left(\tilde{X}\right)$. If $f$ is an increasing function of $x$ between $X_{min} = \min(x)$ and $X_{max} = \max(x)$, $Y_{up}$ and $Y_{low}$ can be defined by $Y_{up} = f\left(X_{up}\right)$, and $Y_{low} = f(X_{low})$. If $f$ is decreasing in the same range, we get $Y_{up} = f(X_{low})$ and $Y_{low} = f\left(X_{up}\right)$.

The upper and lower output uncertainties can be defined by the left and right deviations of $y$ about its best estimate $\tilde{Y}$, so that $\sigma_{Y+} = Y_{up} - \tilde{Y}$, and $\sigma_{Y-} = \tilde{Y} - Y_{low}$.

For small input uncertainties $\sigma_{X\pm}$, one can apply the approximation of a second-order Taylor series expansion to compute the upper/lower output uncertainties $\sigma_{Y\pm}$. Omitting the terms leading to moments higher than the second one, Winzer (2000) gives

$$\sigma_{Y\pm}^2 \approx \left(\left.\frac{\partial f}{\partial x}\right|_{x=\tilde{X}}\right)^2 \sigma_{X\pm}^2 - \frac{1}{4}\left(\left.\frac{\partial^2 f}{\partial x^2}\right|_{x=\tilde{X}} \sigma_{X\pm}^2\right)^2, \quad (A.2)$$

where $\partial f/\partial x$ and $\partial^2 f/\partial x^2$ respectively denote the first and second partial derivatives of $f$ with respect to $x$. If $\sigma_{X+} = \sigma_{X-} = \sigma_X$, Eq. (A.2) is the general law of uncertainty propagation. Throughout our study, we apply the second-order approximation as long as the input uncertainties are less than the arbitrary value 20%. For larger uncertainties, the second-order approximation can introduce bias in the error estimate because of the use of a truncated series expansion, and we compute the output uncertainties thanks to the transformed bounds of the confidence interval.

For practical reasons, it is often more convenient to associate a single uncertainty value $\sigma_X$, hereafter denoted $\sigma$ in order to simplify the notations, rather than dealing with asymmetric uncertainties $\sigma_{X\pm}$, hereafter denoted $\sigma_\pm$. Most people remove the asymmetry by taking the highest value between $\sigma_+$ and $\sigma_-$, or by averaging the two values, arithmetically or geometrically. Although the arithmetic mean gives the correct uncertainty in most cases of practical interest and small uncertainties (D'Agostini & Raso, 2000), we can follow a statistical approach based on asymmetrical probability density functions (pdf), also applicable with large uncertainties.

In the general case where the $100(1 - \alpha)\%$ CI does not encompass the whole distribution of the estimates X of $x$, asymmetric uncertainties need careful handling with known likelihood functions (Barlow, 2003). If the CI bounds $X_{up}$ and $X_{low}$ are close to the extremal values $X_{max}$ and $X_{min}$ of the distribution, and if there is no specific knowledge about the distribution itself, one can use the standard deviation of an asymmetric distribution as estimator for the symmetric uncertainty. When only the value of the best estimate $\tilde{X}$ is known, in addition to the upper and lower bounds of the CI, it is reasonable to assume that the probability to obtain values near the bounds is lower than values near $\tilde{X}$. In this case, a simple approximation of the pdf is given by the asymmetric triangular distribution, with mode $\tilde{X}$, width $X_{max} - X_{min}$, and variance

$$\sigma_{tri}^2 = \frac{(X_{max} - X_{min})^2}{18}\left[1 - \frac{\left(X_{max} - \tilde{X}\right)\left(\tilde{X} - X_{min}\right)}{(X_{max} - X_{min})^2}\right]. \quad (A.3)$$

Kotz & Van Dorp (2004) give the analytic relations

$$X_{max} = \frac{X_{up} - \tilde{X}\sqrt{\frac{\alpha/2}{1-q}}}{1 - \sqrt{\frac{\alpha/2}{1-q}}} = \tilde{X} + \frac{\sigma_+}{1 - \sqrt{\frac{\alpha/2}{1-q}}}, \quad (A.4)$$

$$X_{min} = \frac{X_{low} - \tilde{X}\sqrt{\frac{\alpha/2}{q}}}{1 - \sqrt{\frac{\alpha/2}{q}}} = \tilde{X} - \frac{\sigma_-}{1 - \sqrt{\frac{\alpha/2}{q}}}, \quad (A.5)$$

where $q$ is solution of the equation

$$q = \frac{\sigma_-}{\sigma_+}\left(1 + \frac{\sigma_-}{\sigma_+}\frac{1 - \sqrt{\frac{\alpha/2}{q}}}{1 - \sqrt{\frac{\alpha/2}{1-q}}}\right)^{-1}. \quad (A.6)$$

Throughout our study, we use the standard deviation of the asymmetric triangular distribution as estimator for the symmetric uncertainty.

# Appendix B: De-reddening

If $m_{band}$ is the observed broad-band magnitude, the dereddened magnitude is $m_{band} - A_{band}$, where $A_{band}$ is the interstellar extinction in the band. To calculate the value of the extinction at any wavelength, we use the relation $A_\lambda = (R_\lambda/R_V)\,A_V$, where $R_\lambda = A_\lambda/(A_B - A_V)$ is the ratio of total to selective extinction at any wavelength (Seaton, 1979), and V and B stand for the visible and the blue wavelengths: $\lambda_V = 0.54\,\mu$m, and $\lambda_B = 0.44\,\mu$m (Cardelli et al., 1989; Williams, 1992).

To get the wavelength dependence of the extinction $R_\lambda$ in the IR/optical region, we use the tabular data of the Asiago Database of Photometric Systems available online, following Fitzpatrick (1999), for the case $R_V = 3.1$.

The visual interstellar extinction $A_V$ is calculated thanks to the numerical algorithm of Hakkila et al. (1997), including the studies of Fitzgerald (1968), Neckel & Klare (1980), Berdnikov & Pavlovskaya (1991), Arenou et al. (1992), Chen et al. (1998), and Drimmel & Spergel (2001), plus a sample of studies of high-galactic latitude clouds. The algorithm calculates the three-dimensional visual extinction from inputs of distance, Galactic longitude and latitude. The final estimate of the visual extinction is given by weighted averaging of the individual study values. Since the datasets used in the analyses are not statistically independent, Hakkila suggests to use the simple

**Table B.1.** Total visual extinction of λ Gru as obtained from the relevant studies.

| Study | $A_V$ |
|---|---|
| Fitzgerald (1968) | 0.00(13) |
| Penprase (1992) | 0.00(24) |
| Arenou et al. (1992) | 0.16(15) |
| Chen et al. (1998) | 0.00(17) |
| Drimmel & Spergel (2001) | 0.04(5) |

averaging of the individual study uncertainties as formal extinction uncertainty.

The total visual extinctions of λ Gru are shown in Table B.1 for the relevant studies, with the distance 76 pc, and the Galactic coordinates $l = 2.2153°$, and $b = -53.6743°$. Because the estimate from Arenou et al. (1992) does not agree with the 4 other estimates, we do not use it for averaging. The weighted average estimate of $A_V$ is 0.03, with a mean uncertainty of 0.15.

## Appendix C: Residual bootstrapping

The bootstrap process is based on the idea that if the original data population is a good approximation of the unknown distribution, each sample of the data population closely resembles that of the original data (Babu & Feigelson, 1996). The bootstrap process can be summarized as follows (Palm, 2002; Dogan, 2007): fabricate many "new" data sets by resampling the original data set, then estimate the angular diameter value $\phi$ for each of these "new" data sets to generate a distribution of the angular diameter estimates, and finally use the resulting empirical distribution of the angular diameters to estimate the confidence intervals.

In the direct method, the resampling with replacement is based on the experimental distribution function of the original data. For regression purpose, it is recommended to instead use the residual-based method, implemented as follows.

1. Fit the model to the original measurements $F_i$, with their standard uncertainties $\sigma_i$ ($i = 1, ...N_{meas}$), by minimizing the $\chi^2(\phi)$ function. Call the best-fit angular diameter $\phi_{best}$, and the associated $\chi^2$ minimum value $\chi^2_{min}$.
2. Compute the residuals $r_i$ using $\phi_{best}$. Call $\hat{F}_i = (\phi^2_{best}/4)M_i$ the projection of the best-fit model on the $i$th spectral channel.
   Because the amplitude of each error term $e_i = F_i - \hat{F}_i$ is correlated with the wavelength, we prefer to use the unscaled Pearson residuals $r_i = e_i/\sigma_i$, instead of the error terms themselves.
3. Center the residuals by subtracting the mean of the original residual terms (Friedmann, 1981). Figure C.1 shows an example of the spectral distribution of the centred Pearson residuals and of the corresponding histogram, obtained by fitting the MARCS-model spectrum on the SWS spectrum as described in Sect. 7.5.
4. Resample the centred residuals by drawing randomly with replacement, so that a new residual value is obtained for each measurement (nonparametric bootstrap). Denote $(r_i)_k$ as the resampled normalized residual term for the $k$th data set at the wavelength $\lambda_i$. We have introduced the subscript $k$ because this step and the next two will be repeated many times.
5. Build new data sets $(F_i)_k$ ($k = 1, ...N_{boot}$) from $(F_i)_k = \hat{F}_i + \sigma_i (r_i)_k$.
6. Estimate the model angular diameter $\phi_{best}$ by $\chi^2$ minimization for each fabricated data set.

Repeat steps 3, 4, 5, and 6 many times to obtain a sufficiently large bootstrap sample (e.g. $N_{boot} = 1\,000$). At the end of the process, we have $N_{boot}$ best-fit angular diameter estimates of $\phi_{best}$, as well as $N_{boot}$ associated $\chi^2_{min}$ values. Figure C.2 shows an example of an empirical distribution of the $(\Delta\chi^2)_k = (\chi^2_{min})_k - \chi^2_{min}$ values given by bootstrapping, compared to the $\chi^2$ distribution with 1 degree of freedom. The right and left panels respectively show the cumulative and the ordinary histograms of the residual-bootstrap $\Delta\chi^2$. The central panel shows the $\chi^2$ theoretical quantile-quantile (QQ) plot of $\Delta\chi^2$, where the cube root scaling has been applied for both the order response values (as ordinates), and the $\chi^2$ order statistics medians (as abscissas), as suggested by Chambers et al. (1983). Used as quantile estimators, the order statistics medians are computed according to NIST/SEMATECH[18]. Weak departures from straightness observed on theoretical QQ plots is an indication of the good agreement between the theoretical and the empirical distributions.

Once the $N_{boot}$ $\chi^2$ minimization procedures are terminated, we can estimate the angular diameter confidence interval from the $\Delta\chi^2$ distribution thanks to the simple percentile confidence interval method, easy to implement (Efron & Tibshirani, 1983): for the $1 - \alpha$ confidence level, calculate $\Delta\chi^2_{\alpha/2}$ and $\Delta\chi^2_{1-\alpha/2}$, the percentiles $100(\alpha/2)\%$ and $100(1 - \alpha/2)\%$ relative to the residual-bootstrap distribution of $\Delta\chi^2$; then, among all the bootstrap angular diameters with associated $\Delta\chi^2$ between $\Delta\chi^2_{\alpha/2}$ and $\Delta\chi^2_{1-\alpha/2}$, find the smallest and the greatest bootstrap angular diameter estimates $\phi^{low}_{boot}$ and $\phi^{up}_{boot}$, corresponding to the lower and the upper bounds of the $\Delta\chi^2$ confidence interval.

Because the mean of the distribution of the $N_{boot}$ bootstrap $\chi^2_{min}$ values is not equal to the minimum $\chi^2$ obtained with the original data set, we instead use the bias corrected percentile confidence interval method (Efron & Tibshirani, 1993). In this method, the probabilities $\alpha/2$ and $1 - \alpha/2$ are replaced by $\alpha_1$ and $\alpha_2$, the values of the standard normal cdf for the points $u_1 = 2u_p + u_{\alpha/2}$ and $u_2 = 2u_p + u_{1-\alpha/2}$, where $p$ is the proportion of negative $\Delta\chi^2$ values, and $u_p$ is the $(100p)$th percentile relative to the standard normal cdf.

Formally, if $\Phi$ is the standard normal cdf, $u_p = \Phi^{-1}(p)$, $u_{\alpha/2} = \Phi^{-1}(\alpha/2)$, $u_{1-\alpha/2} = \Phi^{-1}(1 - \alpha/2)$, $\alpha_1 = \Phi(2u_p + u_{\alpha/2})$, and $\alpha_2 = \Phi(2u_p + u_{1-\alpha/2})$.

Finally, $\phi^{low}_{boot}$ and $\phi^{up}_{boot}$ are the smallest and the greatest bootstrap angular diameter estimates with the $\Delta\chi^2$ values between $\Delta\chi^2_{\alpha_1}$ and $\Delta\chi^2_{\alpha_2}$.

**List of Objects**

---

[18] www.itl.nist.gov/div898/handbook/

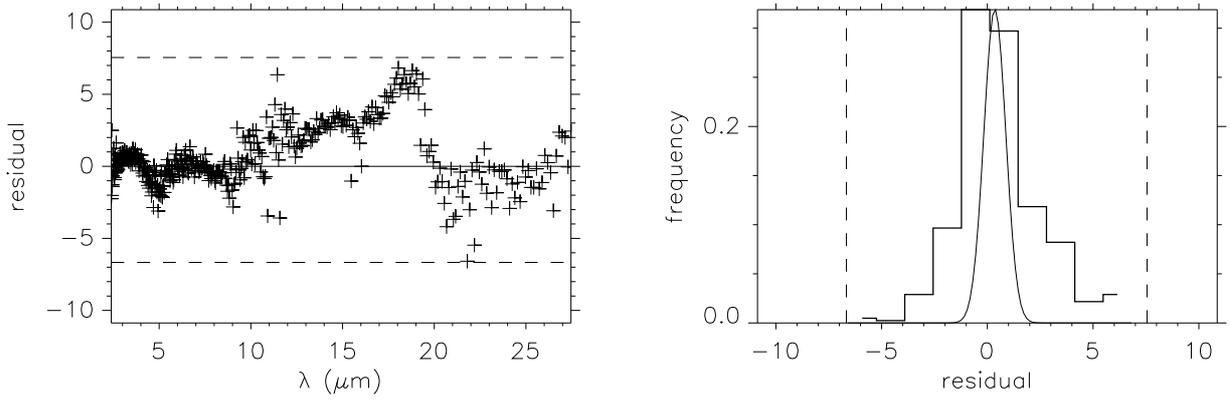

**Fig. C.1.** Left panel: plot of the spectral distribution of the original centred Pearson residuals, obtained by fitting the MARCS model on the SWS spectrum. The residual values located above and below the two horizontal dashed lines are identified as extreme outliers. Right panel: histogram of the Pearson residual distribution. The thin curve within the histogram is the normal probability density function shown for comparison. The two vertical dashed lines give the positions of the upper and lower outer fences identifying the extreme outliers.

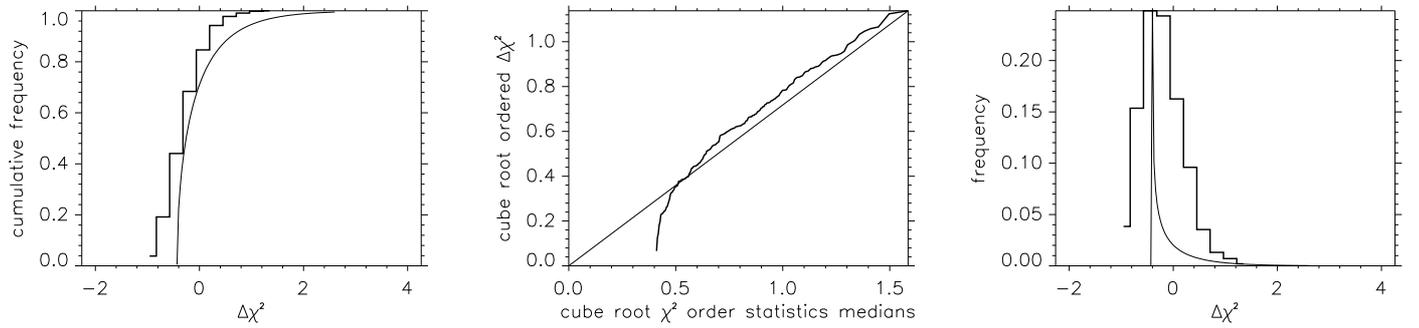

**Fig. C.2.** Left panel: cumulative histogram of the $\Delta\chi^2$ values given by residual bootstrapping. The thin curve shows the $\chi^2$ cumulative distribution function for comparison, with 1 degree of freedom. Central panel: corresponding $\chi^2$ theoretical QQ plot, where the cube root of the ordered $\Delta\chi^2$ values are plotted against the cube root of the $\chi^2$ order statistics medians. The straight line is added for reference. Right panel: histogram of the residual-bootstrap $\Delta\chi^2$, and $\chi^2$ probability density function ($\nu = 1$) for comparison (thin curve).